\documentstyle[aps,epsf]{revtex}

\begin{document}
%\twocolumn
%\draft
\title{Quantized circular motion of a trapped Bose-Einstein condensate:
        coherent rotation and vortices}

\author{Karl-Peter Marzlin and Weiping Zhang}
%\\[2mm]
\address{
School of Mathematics, Physics,
Computing and Electronics, Macquarie University, Sydney, NSW 2109,
Australia
}
\maketitle
%%%%%%%%%%%%%%%%%%%%%%%%%%%%%%%%%%%%%%%%%%%%%%%%%%%%%%%%%%%%%%%%
\begin{abstract}
We study the creation of vortex states in a trapped Bose-Einstein
condensate by a rotating force.
For a harmonic trapping potential the rotating force induces
only a circular motion of the whole condensate
around the trap center which does not depend on the 
interatomic interaction. 
For the creation of a pure vortex state
it is necessary to confine
the atoms in an anharmonic trapping potential.
The efficiency of the creation can be greatly enhanced by
a sinusodial variation of the force's angular velocity.
We present analytical and numerical calculations for the case
of a quartic trapping potential. The physical mechanism behind
the requirement of an anharmonic trapping potential for the
creation of pure vortex states is explained.
\end{abstract}
%%%%%%%%%%%%%%%%%%%%%%%%%%%%%%%%%%%%%%%%%%%%%%%%%%%%%%%%%%%%%%%%
$ $ \\
%\narrowtext
%%%%%%%%%%%%%%%%%%%%%%%%%%%%%%%%%%%%%%%%%%%%%%%%%%%%%%%%%%%%%%%%
\section{Introduction}
The experimental realization of magnetically trapped Bose-Einstein
condensates (BEC) of alkali-metal atoms \cite{experimente} has led to
a broad interest in the properties of this state of quantum matter.
One aspect of particular interest is to study the superfluid behaviour of
these systems. An ideal tool to examine this is to rotate the 
condensate and to observe whether such a rotation can result in
the creation of vortex states. 

Theoretically many aspects of vortex states in a trapped BEC have been
discussed in the literature.
The critical angular velocity required to create a vortex state in trapped 
BECs was derived by Baym and Pethick \cite{baym96}. Dalfovo and Stringari
determined numerically the shape of the vortex state \cite{dalfovo96} and 
the moment of inertia in harmonic traps\cite{stringari96}.
Ho and Shenoy \cite{ho96} discussed the influence of the
internal atomic structure on vortex states and Barenghi \cite{barenghi} 
studied vortex waves. Wilkin {\em et al.} \cite{wilkin97}
considered the fragmentation of 
a rotating BEC with attractive interaction.
Dodd {\em et al.}\cite{dodd97} have numerically studied excitations 
around a vortex state. Rokshar showed that the stability of a vortex state 
depends on the shape of the trap \cite{rokshar97a} and considered the
properties of vortex states on a torus \cite{rokshar97b}.

Despite recent interest in the rotational properties of BECs,  
the experimental realization of a vortex state remains to be done
\cite{ketterlepisa}. In contrast to superfluid Helium, the trapped  
BECs are not in direct contact to an external container. How to rotate
the trapped BECs into vortex states is still an open question. We have recently
proposed to transfer angular momentum 
to a BEC by using multiple travelling wave laser beams \cite{marzlin97a} and 
to create vortex states
by means of Laguerre-Gaussian laser beams using a Raman transition 
\cite{marzlin97b}. A related idea based on Raman transitions 
has also independently been proposed 
by Bolda and Walls \cite{bolda97}. In addition, a combination of multiple laser 
beams with Raman transition has recently been employed to discuss the creation
of dark solitons and vortices in trapped BECs \cite{zoller97}. 

In the present paper we study different aspects of
the quantized circular motion of a trapped BEC 
driven by a spatially homogeneous rotating force.
In Ref.~\cite{marzlin97a} we showed that such a force can be
created by four travelling  
wave laser beams. In particular, we analyse the reason why
in a harmonic trap the rotating force only leads to 
a rotational state of BECs which is a coherent superposition
of vortex states.
On this basis, we further demonstrate how to create a pure vortex state 
by introducing an anharmonic trapping potential. 

The paper is organised as follows. In Sec. II we
derive the Gross-Pitaevskii equation for trapped
BECs composed of two-level atoms in the presence of highly detuned laser 
fields with different frequencies.
In Sec. III, we review the proposal of Ref.~\cite{marzlin97a} and
apply the general model to the transfer of angular momentum
to BECs by a configuration of four travelling-wave laser beams.
The force induced by these laser beams is spatially
homogeneous and rotates with the frequency difference between
the laser beams.
In Sec. IV we discuss the center-of-mass motion and the shape of a
harmonically trapped BEC driven by the rotating force.
Further we analyze the physical mechanism  of why rotating a
harmonically trapped BEC cannot create a pure vortex state.
The possibility to create a pure vortex state of an ideal
BEC is demonstrated in Sec.~V by introducing an anharmonic
trapping potential. In Sec.~VI we consider an interacting
Bose gas in an anharmonic trap. With a sinusodial modulation
of the frequency difference between the laser beams the
nonlinear energy shift due to the interatomic interaction
can be compensated. We numerically show that an almost pure
vortex state can be created. The conlusion is included in
Sec.~VII with a short discussion of
the relation between the present proposal
and that of Refs. \cite{marzlin97b,bolda97} based on Raman transitions.
%%%%%%%%%%%%%%%%%%%%%%%%%%%%%%%%%%%%%%%%%%%%%%%%%%%%%%%%%%%%%%%%
\section{Gross-Pitaevskii equation for a BEC in multiple laser beams}
We consider a condensate of $N$ two-level atoms at zero temperature
confined in a trapping potential. In the formalism of quantum field
theory this system can be described by the Hamiltonian
\begin{equation} 
  H_0 = \int d^3x \sum_{i=e,g} \Big \{ \Psi_i^\dagger ( \vec{x}) 
  \Big [H_{c.m.}  + E_i -\mu \Big ] \Psi_i (\vec{ x}) \Big \}
  +
  {1\over2} \int d^3 x \sum_{i,j=e,g}
  g_{ij}\Psi_i^\dagger (\vec{ x})\Psi_j^\dagger (\vec{x}) 
  \Psi_j(\vec{x}) \Psi_i (\vec{ x})  \; .
\label{h0} \end{equation}
The symbol $\mu$ denotes the chemical potential.
The center-of-mass Hamiltonian
\begin{equation}
  H_{c.m.} := \frac{ -\hbar^2 \Delta }{2M} + V_{\mbox{{\scriptsize trap}}}
  (\vec{x})
\label{hcms} \end{equation} 
contains the kinetic energy and the trapping potential 
$V_{\mbox{{\scriptsize trap}}}(\vec{x})$, which
is assumed to be cylindrically symmetric around the z-axis.
$\Psi_i (\vec{ x}) $ denote the
field operator for atoms in the internal ground state ($i=g$)
and excited state ($i=e$) with internal energy $E_i$. 
$M$ dentoes the atomic mass. 
Without affecting the generality of our final results
we have assumed that the trap potential is the same for both internal
states.
The second integral in Eq. (\ref{h0}) describes the interatomic
interaction in the absence of laser fields.
The factors $g_{ij}$ are defined as $4\pi \hbar^2 a_{ij} /M$, 
with $a_{ij}$ being the s-wave scattering lengths for
atoms with different internal states.

The interaction of the atoms with the laser beams is incorporated
by the electric dipole coupling in the rotating wave approximation,
\begin{equation}
  H_{int} = -\int d^3 x \Big \{ \Psi_e^\dagger (\vec{ x}) \Psi_g
  (\vec{ x}) \sum_{a=1}^n \hbar \Omega_a^{(+)}(\vec{x}) e^{-i\omega _a t}
  + H.c. \Big \} \; .
\end{equation}
Here $\omega _a , a= 1,\ldots ,n$ is the frequency of the $a$th
laser beam, and $\Omega_a^{(+)} (\vec{ x})$ is defined as
$ \vec{d}\cdot \vec{E}_a^{(+)}(\vec{x})/\hbar$ with $\vec{ d}$
being the dipole moment of the atoms and $\vec{ E}_a^{(+)}$
denoting the positive frequency part of the $a$th laser's
electric field.

It has been discussed at length in the literature
\cite{zhang94,lenz94,lewenstein94} how this Hamiltonian can
be reduced to an effective Hamiltonian for atoms being
in their internal ground state in the presence of a single laser beam.
For this reason we will only briefly discuss how this can be
done in the presence of several laser beams with different
frequencies. In order to avoid spontaneous emission all laser
beams have to be detuned far off resonance so that the population of 
the excited-state atoms is negligible. In this case 
the Heisenberg equation of motion for the excited-state field operator can be 
adiabatically eliminated by assuming that the detunings
$\Delta _a := \omega _a -(E_e-E_g)/\hbar$ 
of the laser beams from the atomic resonance frequency
are all in the same
order of magnitude and are the largest characteristic 
frequencies of the system \cite{zhang94}.
The resulting expression for $\Psi_e$ is given by
\begin{equation}
  \Psi_e(\vec{x},t) \approx -\sum_a \frac{\Omega _a^{(+)} e^{-i
  \omega _a t} }{2(\Delta _a + i \gamma /2)} \Psi_g (\vec{x},t)+
  {i\over 2} \int d^3 y L^* (\vec{x}-\vec{ y}) \sum_{a=1}^n
  \frac{\Omega_a^{(+)}(\vec{y}) e^{-i \omega _a t} }{
  (\Delta _a + i \gamma /2)^2} \Psi_g^\dagger (\vec{y},t)
  \Psi_g (\vec{ y},t) \Psi_g (\vec{ x},t) \; .
\label{psie} \end{equation} 
The Kernel $L(\vec{ x}-\vec{ y})$ describing the photon-exchange induced 
dipole-dipole interaction is defined in Ref. \cite{zhang94}.
The parameter $\gamma$ denotes the
sponataneous emission rate from excited atoms and is assumed
to be much smaller than the detuning of the laser beams in this paper.
The expression (\ref{psie}) for $\Psi_e$ can be inserted into
the Heisenberg equation for $\Psi_g$ thus leading to the effective
equation
\begin{equation} 
  i\hbar \dot{\Psi}_g(\vec{x},t) = - [H_0, \Psi_g (\vec{ x},t)] +
  V_L(\vec{x},t) \Psi_g(\vec{x},t) + V_{NL} (\vec{x},t)
  \Psi_g (\vec{x},t) \; ,
\label{qftdgl} \end{equation}
with the light-induced linear potential being given by
\begin{equation}
  V_L(\vec{x},t) := \frac{\hbar}{4(\bar{\Delta}+ i \gamma /2)}
  \sum_{a,b=1}^n \Omega _a^{(+)}(\vec{x}) \Omega _b^{(-)} (\vec{x})
  e^{-i (\omega _a -\omega _b)t} \; .
\label{vl} \end{equation}
Note that in the denominator we have replaced $\Delta _a$
by the average detuning $\bar{\Delta}$ of all lasers. This
has to be done for consistency since otherwise $V_L$ would be
non-Hermitian even in absence of spontaneous emission.
Mathematically this can be made because the difference
$\Delta _b - \Delta _a$ between two detunings has to be treated
as of higher order in $1/ \bar{ \Delta }$. Assuming that the condensate
has a coherence length larger than the characteristic distance 
of the photon-exchange induced dipole-dipole interation, we derive
the nonlinear potential
\begin{equation}
  V_{NL}(\vec{x},t) = g_{eg} \Psi_g^\dagger  (\vec{x},t)
  \Psi_g(\vec{x},t) \sum_{a,b=1}^n \frac{\Omega _a^{(+)}(\vec{x})
  \Omega _b^{(-)}(\vec{x}) e^{-i(\omega _a-\omega _b)t} }{
  4(\bar{\Delta}^2 + \gamma ^2 /4)}
\end{equation}
The strength of this potential depends on the intensity of the applied
laser fields. As we will be concerned only with very weak laser beams
$V_{NL}$ can safely be neglected in the present case.

To describe the collective evolution of a condensed gas of bosonic
atoms we will adopt the mean field approach in Eq. (\ref{qftdgl}).
This approximation describes a trapped Bose-Einstein
condensate quite well \cite{hartreegood} and amounts in
replacing the field operator $\Psi_g(\vec{x},t)$ in Eq. (\ref{qftdgl}),
after the calculation of the commutator, by a collective wavefunction
$\sqrt{N} \psi (\vec{x},t)$ 
of all atoms in the condensate, with
$\psi$ being normalized to one. This procedure
results in the Gross-Pitaevskii equation for $\psi$,
\begin{equation}
  i\hbar \frac{d\psi(\vec{x},t)}{dt} = \{ H_{c.m.}+V_L(\vec{x},t)
  + g N |\psi(\vec{ x},t)|^2 \} \psi(\vec{ x},t) \; .
\label{gpe} \end{equation} 
For brevity we have introduced $g:= g_{gg}$. This equation determines  
the nonlinear dynamics of trapped BECs in multiple laser beams 
and defines the effective Hamiltonian 
\begin{equation}
H = H_{c.m.} + V_L + N g |\psi|^2 \psi
\label{gpeham} \end{equation} 
for the collective wavefunction $\psi$.
%
%%%%%%%%%%%%%%%%%%%%%%%%%%%%%%%%%%%%%%%%%%%%%%%%%%%%%%%%%%%%%%
\section{Transfer of angular momentum by an optical potential}
The design of an optical potential $V_L$ which can rotate the
BEC by transferring angular momentum 
was addressed in Ref. \cite{marzlin97a}
to which we refer for all details of the construction. Here we
will only briefly sketch the important features of this
optical potential.

The potential $V_L(\vec{x},t)$ of Ref. \cite{marzlin97a} is created
by four traveling wave laser beams of equal intensity and width much
larger than the size of the condensate (this is usually the case)
so that they can be considered as plane waves.
The laser beams 1 and 2 propagate along the x- and y-direction have the
same frequency $\omega_1= \omega_2 \equiv \omega $. Their phases are chosen to
be equal at the origin so that the Rabi frequencies
of the two lasers have the form
$\Omega_1^{(+)}(\vec{x}) = \Omega_0 \exp [i k x]$
and $\Omega_2^{(+)} (\vec{x}) = \Omega_0 \exp [i k y] $.
The other two laser beams have the frequency
$\omega_3 = \omega_4 \equiv \omega^\prime$ and propagate in
directions slightly different
from the x- and y- direction with a small angle $\theta$.
The frequency difference $\omega -\omega^\prime$
is assumed to be so small that the difference between $|\vec{k}_a|$
and $|\vec{k}_b|$ for any $a,b,=1,\ldots ,4$ is negligible.
The wavevector of the third laser beam is given by
$k \cos (\theta) \vec{e}_x - k \sin (\theta) \vec{e}_y \approx
k \vec{e}_x - \delta k \vec{e}_y$, where
\begin{equation}
  \delta k := k \theta
\end{equation}
denotes the deviation of the wavevectors.
Similiarly the 4th laser's wavevector is
given by $k \vec{e}_y - \delta k \vec{e}_x$.
The phases are chosen so that laser 4 is in phase with lasers 1 and 2
at the origin whereas the phase of laser 3 is shifted by
$-\pi/2$. The third Rabi frequency is then given by
$\Omega_3^{(+)}(\vec{x}) = \exp [-i \pi/2] \Omega_0 \exp [i (
k x -\delta k y)]$ and the fourth by
$\Omega_4^{(+)}(\vec{x}) = \Omega_0 \exp [i (k y - \delta k x)] $.

The described configuration of laser beams leads to the optical potential
\begin{eqnarray}
V_L(\vec{x},t)&=& \hbar \Omega_{\mbox{{\scriptsize eff}}} \Big \{
  e^{-i k(x-y)} [1+i e^{-i \delta k (x-y)}] 
\nonumber \\ & &
  + e^{-i \phi(t)} \big [ e^{i \delta k x} + i e^{i\delta
  k y} +i e^{-i k(x-y) +i \delta k y}
  + e^{ik(x-y)+i\delta kx}
  \big ] \Big \} +H.c.\; ,
\label{vl2} \end{eqnarray}
where we have introduced the effective Rabi frequency
$\Omega_{\mbox{{\scriptsize eff}}} := |\Omega_0|^2/(4 \bar{\Delta})$
and the time dependent phase $\phi(t) := t(\omega - \omega^\prime)$.
In this section and in the rest of the paper we neglect c-number terms 
in the optical potential since they only produce an overall
phase shift in the wavefunction.

Since the wavelength of an optical laser is much smaller than
the size of a typical condensate (several micrometers) all exponentials
in Eq. (\ref{vl2}) containing the wavevector $k$ are rapidly varying
over the condensate and therefore average to zero \cite{marzlin97a}.
The remaining exponentials, $\exp [i \delta k x]$ and $\exp [i \delta
k y]$, can be approximated by expressions linear in $\delta k$ if
the angle $\theta$ is small enough. The resulting optical potential
then takes the form.
\begin{equation}
  V_L(\vec{x},t) = 2 \hbar \Omega_{\mbox{{\scriptsize eff}}} 
  \delta k \{ \sin (\phi(t))
  x - \cos(\phi(t)) y  \}\; .
\label{vllin} \end{equation}
It obviously produces a rotating spatially homogeneous
force over the size of the condensate. 
The rotating force can transfer angular
momentum to the condensate. This can be seen easily
by rewriting it in the form
\begin{equation}
  V_L(\vec{x},t) = i\hbar \Omega_{\mbox{{\scriptsize eff}}} \delta k
  \{ (x+iy)e^{-i\phi(t)} - (x-i y) e^{i\phi (t)} \}\; .
\label{vlvort} \end{equation}
Because of the commutation relation 
$[L_z ,(x\pm i y)] = \pm \hbar (x\pm iy)$ the operator $x+iy$
increases the orbital angular momentum by $\hbar$. Thus, if the
frequency difference $\omega-\omega^\prime$ is nearly resonant with
a transition between two trap eigenstates, the angular momentum
of the condensate is increased by the force \cite{marzlin97a}.

Since $V_L$ is produced by laser beams we have to estimate the
rate of decoherence of the condensate due to the spontaneous
emission of photons which is given by the probability that
an atom is excited times the spontaneous emission rate $\gamma$.
The excitation probability for off-resonant light is of the order
of $|\Omega_0|^2/\bar{\Delta}^2 = \Omega_{\mbox{{\scriptsize eff}}} /
\bar{\Delta}$. The decoherence time is therefore given by
$\bar{\Delta}/(\gamma \Omega_{\mbox{{\scriptsize eff}}} )$ and is
therefore much larger than $1/\Omega_{\mbox{{\scriptsize eff}}} $
since the detuning is much larger than the natural linewidth.
For instance, if $1/\Omega_{\mbox{{\scriptsize eff}}} )$ is about
10 ms then for $\bar{\Delta} = 100 \gamma$ the decoherence time
is in the order of a second.

One also has to address the question of how stable the frequencies
of the laser beams must be so that $\omega-\omega^\prime$
is well defined. It turns out that this poses no problem at all
since only the frequency {\em difference} $\omega-\omega^\prime$
is required to
be stable. Since in the proposal of Ref. \cite{marzlin97a} all four
laser beams originate from the same laser source, and since the
frequency difference is created by modulating the laser beams, it
is clear that only the modulator frequency must be stable. The
stability requirement on the laser beams can therefore
easily be fulfilled by any common laser source.

We also point out that laser beams are of course only one
possibility to produce a rotating homogeneous force. The same
potential could be created by spatially varying and rotating
magnetic fields, for instance. The following analysis of the
condensate's motion in the potential (\ref{vlvort}) therefore
covers this case as well.
%%%%%%%%%%%%%%%%%%%%%%%%%%%%%%%%%%%%%%%%%%%%%%%%%%%%%%%%%%%%%
\section{Harmonically trapped condensate: coherent rotation}
We now turn to the effect of the optical potential (\ref{vlvort})
on a condensate trapped in a harmonical potential of the form
\begin{equation}
  V_{\mbox{{\scriptsize trap}}}  (\vec{x}) = 
  {M\over2} \omega_z^2 z^2  + {M\over 2} \omega_\perp^2 (x^2+y^2)\; .
\end{equation}
$\omega_z$ and $\omega_\perp$ denote the trap frequency in the
z-direction and in the x-y-plane, respectively.

We have shown in Ref. \cite{marzlin97a} that in the case of an ideal
Bose gas and a resonant optical potential (i.e., $\omega-\omega
^\prime = \omega_\perp$) the condensate rotates around the center
of the trap with increasing mean radius but preserving its shape
(comp. Fig. 1). It was also demonstrated that this motion corresponds
to a coherent superposition of vortex states $\psi_n$ ($\propto \exp [i n
\varphi]$, where $\varphi$ is the azimuthal angle in the x-y-plane)
with different winding number $n$. 

In this section we extend this result to an interacting Bose gas.
We start from Eq. (\ref{gpeham}) including 
the resonant optical potential (\ref{vlvort}).
As is well-known the time evolution of the expectation value of any
operator $O$ is governed by the equation
\begin{equation} 
  \frac{d}{dt} \langle O \rangle  = \frac{i}{\hbar}
  \langle  [H,O] \rangle \; .
\end{equation}
It is straightforward to derive from this relation the corresponding
equations for the position operators. Observing that $\int d^3 x
|\psi|^2 \partial_i |\psi|^2 $, $i=x,y,z$ vanishes the center-of-mass
motion takes the simple form
\begin{equation}
  \frac{d^2}{dt^2} \langle \xi \rangle + \omega_\xi^2
   \langle \xi\rangle  = f_\xi (t)\; ,
\end{equation}
for ($\xi = x,y,z$), where the light-induced force $\vec{f}(t)$ is given by
$\vec{f}(t) = 2 (\hbar \Omega_{\mbox{{\scriptsize eff}}} \delta k /M)
  \{ -\sin (\omega_\perp t)
  \vec{e}_x + \cos (\omega_\perp t) \vec{e}_y \}$. For a 
BEC in the ground state the
condensate's mean position and mean momentum are initially zero
($\langle \vec{x} \rangle (0) =0\; ,\; \langle \vec{p}\rangle (0) = 0$)
so that we find for the solution
\begin{equation}
  \langle \vec{x} \rangle = R_0 \{ \vec{e}_x [\omega_\perp t \cos 
  (\omega_\perp t) - \sin (\omega_\perp t)] + \vec{e}_y
  \omega_\perp t \sin (\omega_\perp t) \} \; ,
\label{trajeclsg} \end{equation}
with $R_0 := \hbar \Omega_{\mbox{{\scriptsize eff}}} \delta k /(M
\omega_\perp^2)$. Again the condensate performs a rotation around the
trap center whereby the radius of the rotation increases linearly
in time. This result was
to be expected since usual interatomic potentials are
translation-invariant and therefore affect the center-of-mass
motion only indirectly. It provides an extension of Ehrenfest's
theorem for interacting Bose condensates.
We remark that the result (\ref{trajeclsg}) deviates from the
corresponding result of Ref.~\cite{marzlin97a} by the factor
$-\sin (\omega_\perp t)$
inside the square brackets. This difference arises because
in Ref.~\cite{marzlin97a} a rotating wave approximation (RWA) 
has been made. As usual for RWA the difference is negligible
for large times $\omega_\perp t \gg 1$.

To demonstrate that, as in the noninteracting case, the shape of the
condensate is not affected by the optical potential we consider the
time evolution of the variance of the position operators. Using
$\int d^3 x \xi |\psi|^2 \partial_\xi |\psi|^2 =-\int d^3 x|\psi|^4/2$
and defining $\Lambda_\xi := \langle \xi p_\xi + p_\xi \xi \rangle 
-2 \langle \xi \rangle \; \langle p_\xi \rangle $ we arrive at
\begin{eqnarray}.
  \frac{d (\Delta \xi)^2}{dt} &=& {\Lambda_\xi \over M} \nonumber \\
  \frac{d \Lambda_\xi}{dt} &=& {2\over M} (\Delta p_\xi)^2 -2 M
  \omega_\xi^2 (\Delta \xi)^2 + g \int d^3 u |\psi (\vec{u})|^4
  \label{varianz} \\
  \frac{d (\Delta p_\xi)^2}{dt} &=& -M \omega_\xi^2 \Lambda_\xi -i
  \hbar g \int d^3 u \{ (\partial_\xi \psi^*(\vec{u}))^2 \psi^2(\vec{u})
   - H.c. \}\; .
\nonumber \end{eqnarray}
Because neither the center-of-mass motion nor the optical potential
enter into these equations we can infer that the shape of the condensate
remains unaffected by the optical potential.

It is worth noting that one can derive from Eq. (\ref{varianz}) the
mean energy conservation law
\begin{equation}
  {1\over 2M}(\Delta \vec{p})^2 + {1\over2} M \{ \omega_\perp^2 
  [(\Delta x)^2 + (\Delta 
   y)^2 ] + \omega_z^2 (\Delta z)^2\} + g \int d^3 u |\psi
   (\vec{u})|^4 = \mbox{ const. } \; ,
\end{equation}
which nicely demonstrates the relation between the interaction energy
and the widths of the condensate's wavefunction.

We conclude that, albeit the potential
(\ref{vlvort}) seems to be perfectly suited for this task, neither
a pure vortex state nor even a vortex line far away from the
center of the trap is created. However,
this does not rule out the possibility to form a vortex
state during the formation of the condensate by directly cooling the 
gas in a rotating trapping potential.
Our result may also be helpful to understand why the recent experiments 
at MIT \cite{ketterlepisa} attempting to observe persistent flows
around a potential wall at the trap center by rotating the trap
cannot achieve the expected result. 

To understand why the potential (\ref{vlvort}) produces a superposition
of vortex states rather than a pure one we observe that it couples only
neighboring eigenstates of the trap. This can be seen by expressing
the coordinates $x,y$ in terms of the corresponding annihilation operators
of the trap (i.e., $x \propto a_x + a_x^\dagger$, for instance).
Since the potential is linear in these operators it couples states
with energy $n \hbar \omega_\perp$ (in the noninteracting case)
to states with energy
$(n+1)\hbar \omega_\perp$. If the coupling is resonant ($\phi(t) =
\omega_\perp t$) the potential (\ref{vlvort}) not only resonantly
transforms the ground state to the first excited vortex state,
but it also resonantly transfers the $n$th vortex state (with
energy $n \hbar \omega_\perp$) to the $(n+1)$th vortex state, because
of the equal energy interval between these energy eigenstates.
As a result, the long-time evolution of the system driven by the 
rotating force will lead to a superposition of vortex states
with different winding number.

This implies that to create a pure vortex state one must destroy 
the equal interval between the trap energy levels to avoid the simultaneous
excitations of different vortex states. We will analyze this in the
next section
%
%%%%%%%%%%%%%%%%%%%%%%%%%%%%%%%%%%%%%%%%%%%%%%%%%%%%%%%%%%%%%%%%%%
\section{Creation of a vortex state in an ideal Bose gas}
In the preceding sections we have presented a scheme to transfer a
harmonically
trapped condensate into a rotating state which is a coherent superposition 
of vortex states. We also analysed why the scheme 
does not create a pure vortex state if the trapping potential is harmonic:
the equal interval between the energy levels in a harmonic trap
causes the condensate to climb up a ladder of resonant transitions
into different vortex states.

The situation is different if an anharmonic (yet still rotationally
symmetric) potential $V_{\mbox{{\scriptsize trap}}}$ is used
in Eq. (\ref{gpe}) for which the interval between the
energy levels is not equal. Therefore, if
the frequency $\omega-\omega^\prime $ in the optical potential
(\ref{vlvort}) is choosen to be in resonance with the
transition from the ground state to the first excited vortex state,
the transition to the next vortex state will in general be out of
resonance (see Fig.~2). This fact can be exploited to prepare a 
pure vortex state in an anharmonic trap.

To estimate this quantitatively we first consider the ideal Bose gas and
expand the wavevector in terms of the simultaneous eigenstates
$\psi_{m,s} $ of the Hamiltonian (\ref{hcms}) and the orbital
angular momentum $J_z$,
$\psi  = \sum_{m,s} \exp [-i m \phi(t)]
c_{m,s}(t) \psi_{m,s}$. 
The eigenstates fulfill the relations $H_{c.m.} \psi_{m,s}  =
E_{m,s} \psi_{m,s} $ and $J_z \psi_{m,s} = m \hbar \psi_{m,s} $.
It should be noted that the inclusion of
the phase factor $\exp [-i m \phi (t)]$ in this expansion is equivalent
to the description of the system in a frame of reference rotating with
angular velocity $\dot{\phi}(t) = \omega -\omega^\prime$ (where we 
left open the possibility that the frequencies can vary in time).
In this description the broken symmetry between states of positive ($m>0$)
and negative ($m<0$) angular momentum becomes obvious.
Since the latter are
counterrotating for $\omega-\omega^\prime >0$ transitions to
these states are highly suppressed.

Using this expansion the Schr\"odinger equation (\ref{gpe}) for the ideal
condensate with $g=0$ can be reduced to
\begin{equation}  i\hbar \dot{c}_{m,s} = \{ E_{m,s}-m \hbar \dot{\phi}(t)
  \} c_{m,s} + \hbar 
  \Omega_{\mbox{{\scriptsize eff}}} \delta k \sum_{s^\prime } \big \{
  q_{m;s,s^\prime} c_{m-1,s^\prime } + 
  q^*_{m+1; s^\prime ,s} c_{m+1,s^\prime } \big \}
\label{crea1} \end{equation} 
with $q_{m;s,s^\prime } := \int d^2 x \psi_{m,s}^*( x+i y)
\psi_{m-1,s^\prime}$.

We assume that, as it is the case for the harmonic potential, $H_{c.m.}$
has a unique ground state with energy $E_{0,0}$ and that the energy
levels $E_{1,s}$ are well separated, with $E_{1,0}$ being the lowest
vortex excitation. In this case, if we choose $\omega-\omega^\prime 
= (E_{1,0}-E_{0,0})/\hbar$ the transition from $\psi_{0,0}$ to
$\psi_{1,0}$ is in resonance while the transition from $\psi_{1,0}$
to $\psi_{2,s}$ is detuned by the amount $\Delta_s := 
\omega-\omega^\prime - (E_{2,s}-E_{1,0}) = E_{2,s}+E_{0,0} - 2 E_{1,0}$.
Furthermore, if the condition
$|\Omega_{\mbox{{\scriptsize eff}}}\delta k q_{2;s,0}|
\ll \Delta_s$ is fulfilled, the transition from
$\psi_{1,0} $ to $\psi_{2,s}$ will be highly supressed so that we
effectively have a two-level system consisting of the states 
$\psi_{0,0}$ and $\psi_{1,0} $. It is obvious that after a time
$\pi/\nu$ the complete population will have been transferred to the first 
excited vortex state, where we have defined the Rabi frequency 
\begin{equation} 
  \nu := |\Omega_{\mbox{{\scriptsize eff}}} \delta k q_{1;0,0}|
\end{equation} 
of the transition between $\psi_{0,0}$ and $\psi_{1,0}$
This situation is mathematically very similar to the resonant excitation
of the first excited state in an atom by a laser beam. In this case
one can safely neglect all states beside the ground and the first
excited state, too.

As a specific example we consider a two-dimensional anharmonic oscillator
for which the trapping potential is given by 
$V_{\mbox{{\scriptsize trap}}}= \kappa r^4$, where
$r$ is the two-dimensional radial variable and $\kappa$
determines the strength of the potential. The neglection of the
z-direction does not alter the predictions for the ideal Bose gas and
is justified for the interacting Bose gas if excitations along the
z-axis are sufficiently supressed.
Approximate eigenstates of the resulting Hamiltonian can be obtained
by application of Ritz's variational method. For later use we include
the nonlinearity and minimize the expression
\begin{equation} 
E = \int d^2x \left \{ \psi^* \left [ \frac{\vec{p}^2}{2M}
    + \kappa r^4 \right ] \psi + \frac{g_{2D}}{2} |\psi|^4 \right \}
\end{equation} 
for the conserved energy. The
coupling constant $g_{2D}$ that we use for an effective
description of the vortex creation in two dimensions is related to the
three-dimensional coupling parameter $g$ by $g_{2D} = g N/ l_0$.
Using the ansatz
\begin{equation}  \psi_{n,0}(\vec{x})  = 
  \sqrt{\frac{\alpha_n^{n+1}}{\pi n!}}
  r^n e^{i n \varphi} e^{- \alpha_n r^2/2}\; .
\label{vari} \end{equation} 
we find for the variational energy eigenvalues the expression
$E_{n,0} = E^{(L)}_{n,0} + E^{(NL)}_{n,0}$, with
\begin{equation} 
  E^{(L)}_{n,0} = {\cal E} {(n+1)\over 2} ( 3 + {\cal G}f_n )
  \left [ \frac{2(n+2)}{1+{\cal G} f_n} \right ]^{1/3} 
  \quad , \quad 
  E^{(NL)}_{n,0} = {\cal E} {\cal G}f_n  (n+1)
  \left [ \frac{2(n+2)}{1+{\cal G} f_n} \right ]^{1/3}
  \; .
\end{equation} 
$E^{(L)}_{n,0}$ represents the contribution of the linear part of the
Hamiltonian (kinetic and potential energy) and $ E^{(NL)}_{n,0}$
denoted the contribution of the nonlinear interaction energy.
We have introduced the energy scale
${\cal E} := \kappa l_0^4$, the dimensionless
interaction parameter
${\cal G} := M g_{2D}/(2\pi \hbar^2)$, and the numerical factor $f_n :=
(2n-1)!!/[(n+1)! 2^n]$. The characteristic length scale of the system
is given by $l_0 := [\hbar^2/(2M \kappa )]^{1/6}$.
For the ideal Bose gas we set ${\cal G}=0$ and find for the lowest
energy eigenvalues the values
$E_{0,0} \approx 2.38\cdot {\cal E}$, $E_{1,0}\approx 5.45 \cdot{\cal E}$,
$E_{2,0}\approx 9 \cdot{\cal E}$, and $E_{0,1}\approx 9.64\cdot {\cal E}$.
We have determined the energy $E_{0,1}$ by using the ansatz
$\psi_{0,1} \propto(\varepsilon^2 -r^2) e^{-\delta r^2/2}$
for the wavefunction. The largeness of $E_{0,1}$ allows us to
neglect all transitions to this state. 

The interaction matrix elements are of the order of $l_0$ for these four
states, but we will need only the value for $q_{1;0,0} \approx 0.79 \cdot
l_0$ and $q_{2;0,0}\approx 1.05\cdot l_0$. The transition from $|0,0
\rangle $ to $|1,0 \rangle $ will be in resonance if we choose the
frequency difference $\dot{\phi} = 3.07\cdot {\cal E}/\hbar$. The
transition from $|1,0 \rangle $ to $|2,0 \rangle $ is then supressed
if the interaction matrix element $\hbar \Omega_
{\mbox{{\scriptsize eff}}}\delta k \cdot q_{2;0,0}$
is much smaller than the energy difference $E_{2,0}-E_{1,0}-\hbar
\dot{\phi} \approx 0.48 \cdot {\cal E}$. This is guaranteed if we
set the transition energy $E_{\mbox{{\scriptsize trans}}} :=\hbar \nu$
equal to ${\cal E}/20$.
The laser scheme presented above then will create a
vortex state after a time $\pi/\nu$. 
We have depicted this physical situation (to scale) in Fig.~3.

To estimate its experimental realizability and to compare it with
the interacting Bose gas we consider the case that the
numerical value of the trapping potential strength $\kappa$ is given
by $7.66\cdot 10^{-8}$ J/m$^4$ so that for $^{87}$Rb atoms 
($M=1.45\cdot 10^{-25}$ kg) the trap size turns out to be
$l_0 = 8.91\cdot 10^{-7}$m (These numerical values turn out to be
convenient for the numerical calculations presented below). 
We then find for the energy scale 
${\cal E} \approx 4.83\cdot 10^{-32}$ J and for the
characteristic time for vortex creation 
$\pi/\nu \approx
70$ ms. This well within the range of current experiments with
Bose-Einstein condensates. We mention that we have also made a
numerical study of this creation process for the numbers given
here. The result is in excellent agreement with our analytical
calculations but will not be presented here since the analytical
results already perfectly describe the physical situation for a
non-interacting Bose gas. For all details of the numerical procedure
we refer to the next section where it is applied to the correponding
case of an interacting Bose gas.
%%%%%%%%%%%%%%%%%%%%%%%%%%%%%%%%%%%%%%%%%%%%%%%%%%%%%%%%%%%%%%%%%%
\section{Creation of a vortex state in an interacting Bose gas}
To study the influence of the atomic interaction analytically
we assume that the
nonlinearity in Eq. (\ref{gpe}) is weak enough so that we still can
use the variational ansatz (\ref{vari}). This is not allowed for the
strong interaction regime where this ansatz becomes meaningless, but
it includes the case that the interaction energy
is of the order of the kinetic and potential energy and thus overcomes
the restriction of perturbation theory.

We further assume that radial excitations of the condensate are
unimportant because the energy levels $E_{m,s}$ ($s>0$) are considerably
larger than the corresponding lowest value $E_{m,0}$ for the same
angular momentum $m\hbar$. We then only need to take into account
the states $\psi_{m,0}$. 
The numerical results presented below
further justify this assumption.
Writing the wavefunction as $\psi = \sum_n c_n(t) \exp[-i n \Delta
\omega t] \psi_{n,0}(\vec{ x})$ and projecting Eq. (\ref{gpe}) onto
$\psi_{m,0}$ we arrive at
\begin{eqnarray} 
i \hbar \dot{c}_m &=& \{ E^{(L)}_{m,0}- m \hbar \dot{\phi} \}c_m +
  \hbar \Omega_{\mbox{{\scriptsize eff}}} \delta k\{ q_{m;0,0} c_{m-1}
  + q^*_{m+1;0,0} c_{m+1} \} 
  \nonumber \\ & &
  + 2 E^{(NL)}_{m,0} |c_m|^2 c_m 
  + \sum_{k\neq m} E^{(S)}_{m,k} |c_k|^2 c_m
  + \sum_{l\neq m}\sum_{k\neq l} c^*_k c_l c_{m+k-l} E^{(AS)}_{m;k,l} \; .
\label{ff} \end{eqnarray} 
Here we have introduced the energies $E^{(S)}_{m,k} = E^{(S)}_{k,m} 
:= 2 g_{2D} \int d^2 x |\psi_{m,0}|^2 |\psi_{k,0}|^2 $ for scattering
between atoms in state $k$ and $m$, and
$E^{(AS)}_{m;k,l} := g_{2D} \int d^2 x
\psi^*_{m,0} \psi^*_{k,0} \psi_{l,0} \psi_{m+k-l,0}$ for anomalous
scattering of atoms. 

To solve Eq. (\ref{ff}) analytically we first make the assumption
that only the two states $\psi_{0,0}$ and $\psi_{1,0}$ are populated.
Though this is not justified for a quantitative analysis
it provides very useful physical insight into the effectiveness
of the process of vortex creation. Eq. (\ref{ff}) then reduces to
\begin{eqnarray} 
i \hbar \dot{c}_0 &=& \left \{ E^{(L)}_{0,0} + 2 E^{(NL)}_{0,0}
  |c_0|^2 + E^{(S)}_{0,1} |c_1|^2 \right \}c_0 +
  \hbar \Omega_{\mbox{{\scriptsize eff}}} \delta k q^*_{1;0,0} c_{1}  
  \label{2levnl1} \\
i \hbar \dot{c}_1 &=& \left \{ E^{(L)}_{1,0} + 2 E^{(NL)}_{1,0}
  |c_1|^2 + E^{(S)}_{0,1} |c_0|^2 - \hbar \dot{\phi} \right \}c_1 +
  \hbar \Omega_{\mbox{{\scriptsize eff}}} \delta k q_{1;0,0} c_{0}\; .  
\label{2levnl2} \end{eqnarray} 
This form of the nonlinear Schr\"odinger equation is particularly 
suited to study the effect of a time-dependent variation of the laser
frequencies $\dot{\phi} = \omega-\omega^\prime$. It was originally
proposed in Ref. \cite{marzlin97b} to use such a variation to
compensate for the nonlinear interaction energy between the atoms.
Here we will go into more detail and examine the question of which
variation is the most effective one to create a vortex state.

Let us first assume that $\dot{\phi}$ is constant and chosen so that
initially ($|c_0|^2 = 1\; , \; |c_1|^2 =0$) the transition from
$\psi_{0,0}$ to $\psi_{1,0}$ is on resonance, i.e.,
\begin{equation}
\dot{\phi} = E^{(L)}_{1,0} - E^{(L)}_{0,0} + E^{(S)}_{1,0} - 
    2 E^{(NL)}_{0,0} \; .
\end{equation}
After some time the population of the two states will be altered
($|c_0|^2 <1 \; , \; |c_1|^2 > 0$ ) so that the magnitude of
the terms in curly brackets in Eqs. (\ref{2levnl1}, \ref{2levnl2})
is changed. Thus, the 2-level system is driven out of resonance
by the nonlinear atomic interaction and a complete transfer of
the condensate from $\psi_{0,0}$ to $\psi_{1,0}$ becomes impossible.

It is not hard to see that a time-dependent variation of $\dot{\phi}$
can compensate for this interaction-induced shift of the resonance
frequency. In Ref. \cite{marzlin97b} we proposed to vary 
$\omega-\omega^\prime$ linearly in time. While this works well
if the rate of change is suitably chosen, it has the disadvantage that
one needs a relatively long time to create a vortex state. This is similar
to a Landau-Zener transition \cite{landau32,zener32} which is destroyed
by nonadiabatic transitions if the frequency's rate of change
is too high (see Ref. \cite{marzlin96}, for instance). 
To improve this situation we assume that $\dot{\phi}$ is implicitely
chosen so that the transition between $\psi_{0,0}$ and $\psi_{1,0}$
is {\em always} on resonance,
\begin{equation}
\dot{\phi} = E^{(L)}_{1,0} - E^{(L)}_{0,0} + E^{(S)}_{1,0} 
    (|c_0|^2-|c_1|^2)+ 2 E^{(NL)}_{1,0} |c_1|^2 - 2 E^{(NL)}_{0,0} 
    |c_0|^2   \; .
\label{rescond} \end{equation}
Under this condition the solution of Eqs. (\ref{2levnl1}, \ref{2levnl2})
is given by 
\begin{eqnarray}
  c_0(t) &=& e^{-i \theta (t)} \cos ( \nu t)
  \nonumber \\ 
  c_1(t) &=& \frac{-i q_{1;0,0} }{ |q_{1;0,0}| } e^{-i \theta (t)} \sin (
  \nu t)\; , 
\label{2levlsg} \end{eqnarray}
where
the phase factor $\theta(t)$ is easy to calculate but not needed here.
This solution corresponds to a resonant Rabi Oscillation between the
two states. Hence the two-level model predicts that
the creation of a vortex state can be as effective as in the linear case
provided the nonlinear interaction energy is compensated for.
The corresponding variation of the frequency difference $\dot{\phi}$
can be found by reinserting Eq. (\ref{2levlsg}) into Eq.
(\ref{rescond}) and turns out to be a simple sinusodial variation
with frequency $2 \nu$. Such a sinusodial variation of the frequency
difference $\dot{\phi}$ can be realized by simple frequency
modulation techniques.

The above analysis was made under the oversimplified assumption that
we only need two levels to take into account. The influence of other
modes can be studied in several ways. For instance, they can be treated
as an elementary excitation around the solution (\ref{2levlsg}).
For the sake of brevity we will not discuss the analytical
treatment here
and present instead numerical solutions of the full nonlinear
Schr\"odinger equation (\ref{gpe}). 

{\em Numerical solution:} Our numerical procedure started
from the ground state of the interacting condensate in the quartic
anharmonic potential. This state was found by propagating a Gaussian
trial wavefunction in imaginary time and normalizing it between each step.
The formation of a vortex state was studied by using the modified
split-step method presented in Ref. \cite{devries86}. This consists in
approximating the time evolution operator $\exp [ -i \Delta t (T+V)/\hbar]$
for small $\Delta t$ 
as $\exp [ -i \Delta t V)/(2\hbar)]\exp [ -i \Delta t T\hbar]
\exp [ -i \Delta t V/(2 \hbar)]$, where $T$ and $V$ denote the kinetic
and potential part of the Hamiltonian, respectively.
The terms containing $V$ are diagonal in position space and can easily
be applied to the wavefunction. The term involving $T$ is computed by
transforming the wavefunction to momentum space by using the
fast-Fourier-transformation algorithm (FFT), applying the operator,
and performing the inverse FFT.

The physical parameters for the trap and the Rb atoms are the same
as  given at the end of the previous section (trap strength $\kappa 
=7.66\cdot 10^{-8}$ J/m$^4$, trap size $l_0 = 8.91\cdot 10^{-7}$ m,
mass $M=1.45\cdot 10^{-25}$ kg). In addition, the
two-dimensional nonlinearity parameter $g_{2D}$ was assumed to be
$2.6 \cdot 10^{-43}$ J m$^2$. For Rb atoms the scattering length
is about 5 nm. Therefore, if we relate an effective 3-dimensional
peak density $\hat{\rho}_{3D}$ to the peak of the 2-dimensional
wavefunction $\hat{\psi}_{2D}$ by $|\hat{\psi}_{2D}|^2 g_{2D} 
= \hat{\rho}_{3D} g$ we can infer that this choice of $g_{2D}$
corresponds to a peak density $\hat{\rho}_{3D}$
of about $2\cdot 10^{13}$ cm$^{-3}$.
The strength of the optical potential (\ref{vlvort}) was determined
by setting $\hbar \Omega_{\mbox{{\scriptsize eff}}} \delta k =
2\cdot 10^{-27}$ J/m. It turned out that for this system a grid
size of $64\times 64$ points was accurate enough for the numerical study
of the creation process.

The most laborious part of the numerical simulation was to find
the proper variation of the frequency difference $\dot{\phi}$.
It is not difficult to find an approximation of the
physical parameters in Eq. (\ref{rescond}) by use of the variational
wavefunctions (\ref{vari}). However, in order to get satisfactory
results, these parameters must be determined more precise than
the variationally calculated values which deviate from the exact
values by up to
10\%. We finally found that the best results are produced
if the sinusodial variation is delayed for a time $t_0=50$ ms 
to compensate for effects of other modes than 
$\psi_{0,0}$ and $\psi_{1,0}$. Thus the frequency difference was
given by  
\begin{equation}
  \dot{\phi} = \left \{
  \begin{array}{cc} \omega_{\mbox{{\scriptsize max}}}
   & t \leq t_0 \\
   {1\over 2} \{ (\omega_{\mbox{{\scriptsize max}}}  +
        \omega_{\mbox{{\scriptsize min}}})
      + (\omega_{\mbox{{\scriptsize max}}}  - 
         \omega_{\mbox{{\scriptsize min}}}  )
   \cos (\mu (t-t_0)) \}
   & t\geq t_0  \end{array} 
  \right . \; ,
\end{equation}
with $\omega_{\mbox{{\scriptsize max}}}  = 1430$ s$^{-1}$,
$\omega_{\mbox{{\scriptsize min}}}  = 1205$ s$^{-1}$, and $\mu =
19.5$ s$^{-1}$. 

The numerical results for this choice of parameters
are shown in Fig.~4 and 5. Figs.~4a-d
show the modulus squared of the wavefunction at different times
of the evolution (0 ms, 87 ms, 142 ms, 247 ms). It shows that
for a moderately interacting Bose gas in an anharmonic trap an
almost pure vortex state can be created. Fig.~5 displays the phase
of the wavefunction at the same times (the coarse-grained structure
at the border of the figures represents numerical noise).
Fig.~5b demonstrates that during the evolution two vortex lines are
created, one of which finally becomes the vortex state in Fig.4d
and the other slips out of the trap. This situation
can be described reasonably well by assuming that the wavefunction
is a superposition of the form $\psi = c_0 \psi_{0,0} + c_1
\psi_{1,0} + c_2 \psi_{2,0}$, where $\psi_{2,0}$ describes a vortex
state with angular momentum $2\hbar$. Fig.~4b shows the corresponding
deformation of the wavefunction. The modulus of the expansion
coefficients $c_i$ can be approximately determined by looking at
Fig.~5 since the closer one (two) vortex lines are to the origin the
larger the value of $c_1$ ($c_2$) is. 
We also mention that the vortex lines
rotate around the origin during the creation process. This can be
seen in an animation of the creation process \cite{movie}. 
%%%%%%%%%%%%%%%%%%%%%%%%%%%%%%%%%%%%%%%%%%%%%%%%%%%%%%%%%%%%%%%%%
\section{Conclusion}
We have presented a scheme based on four laser beams 
which induces a spatially homogeneous force to transfer
angular momentum to a trapped condensate. 
For a harmonic
trap this scheme is equivalent to a rotation of the trapping
potential around its original position. In this case the condensate
will be prepared into a superposition of pure vortex states,
corresponding to a circular motion of the condensate as a whole
around the original trap center. The reason for the creation of a
superposition rather than a pure vortex state is that the energy
eigenvalues of the trap are equidistant for a harmonic trap. Thus,
transitions to neighboring states are simultaneously resonant.
In an anharmonic trap this phenomenon does not occur.
If the intensity of the laser
beam is sufficiently low one can tune the transition from the ground
state to the first excited vortex state into resonance while leaving
the transition to the second excited vortex state far off-resonant.
The interaction-induced energy shift appearing in a non-ideal BEC
can be compensated by a sinusodial variation of the rotation frequency
of the force.
It then is possible to create a pure vortex state in an anharmonic trap.

Beside employing an anharmonic trap one alternative method to create
a vortex state is to use Raman transitions induced by
Laguerre-Gaussian laser beams, as it was propsed in Refs.
\cite{marzlin97b,bolda97}. In this case a single Laguerre-Gaussian laser beam
already carries an orbital angular momentum which can be transferred
to the condensate. It is of interest to understand why the proposal
with Raman transitions also works for a harmonic trapping potential
and is not restricted to anharmonic potentials.
In the notation used in the
present paper the basic mechanism of the scheme based on Raman-transitions
is that laser beams of the Laguerre-Gaussian type couple the
condensate, which is initially in the ground state $|0,0 \rangle $ of
the trap and the internal state $| - \rangle $, to a vortex state $
 |1,0 \rangle $ and the internal state $|+ \rangle $.
Here the state vectors are defined by $\psi_{m,s}(\vec{x})
 = \langle \vec{x} | m,s\rangle$.
The interaction Hamiltonian is then roughly of the form $V_L\propto
|1,0 \rangle \otimes |+ \rangle \langle  - | \otimes \langle 0,0|
+ |0,0 \rangle \otimes  |- \rangle \langle + | \otimes \langle 1,0|$.
This combination of internal transition and excitation of the
center-of-mass motion guarantees that the condensate never can occupy
higher vortex states like $|2,0 \rangle \otimes  |\pm \rangle $ since the
condensate can only oscillate between the states 
$|0,0 \rangle \otimes  |- \rangle $ and $|1,0 \rangle \otimes  
|+ \rangle $ in the Raman coupling. 
Hence selective excitations of the internal state by Raman transitions
avoid the resonant excitation of the ladder of transitions to
higher vortex states.
%%%%%%%%%%%%%%%%%%%%%%%%%%%%%%%%%%%%%%%%%%%%%%%%%%%%%%%%%%%%%%

{\bf Acknowledgement}: We thank E.M. Wright and E. Bolda for
helpful discussions. The work has been supported by the Australian
Research Council.
%%%%%%%%%%%%%%%%%%%%%%%%%%%%%%%%%%%%%%%%%%%%%%%%%%%%%%%%%%%%%%%%
%\newpage

%%%%%%%%%%%%%%%%%%%%%%%%%%%%%%%%%%%%%%%%%%%%%%%%%%%%%%%%%%%%%%%%%%%%
\newpage
\begin{figure}[t]
\epsfysize=7cm
\hspace{3cm}
\epsffile{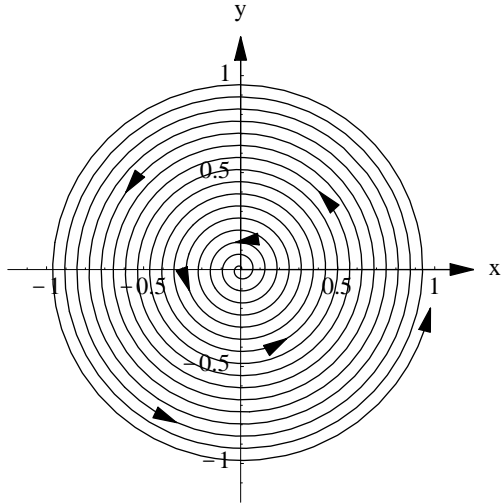}%\vspace{-1cm}
\caption{The trajectory of the center-of-mass of a harmonically trapped
condensate under the influence of a rotating homogeneous force.
x and y are given in units of the trap size $R_\perp$.
During the rotation the shape of the condensate is preserved. }
\end{figure}

$ $

$ $

\begin{figure}[t]
\epsfysize=7cm
\hspace{3cm}
\epsffile{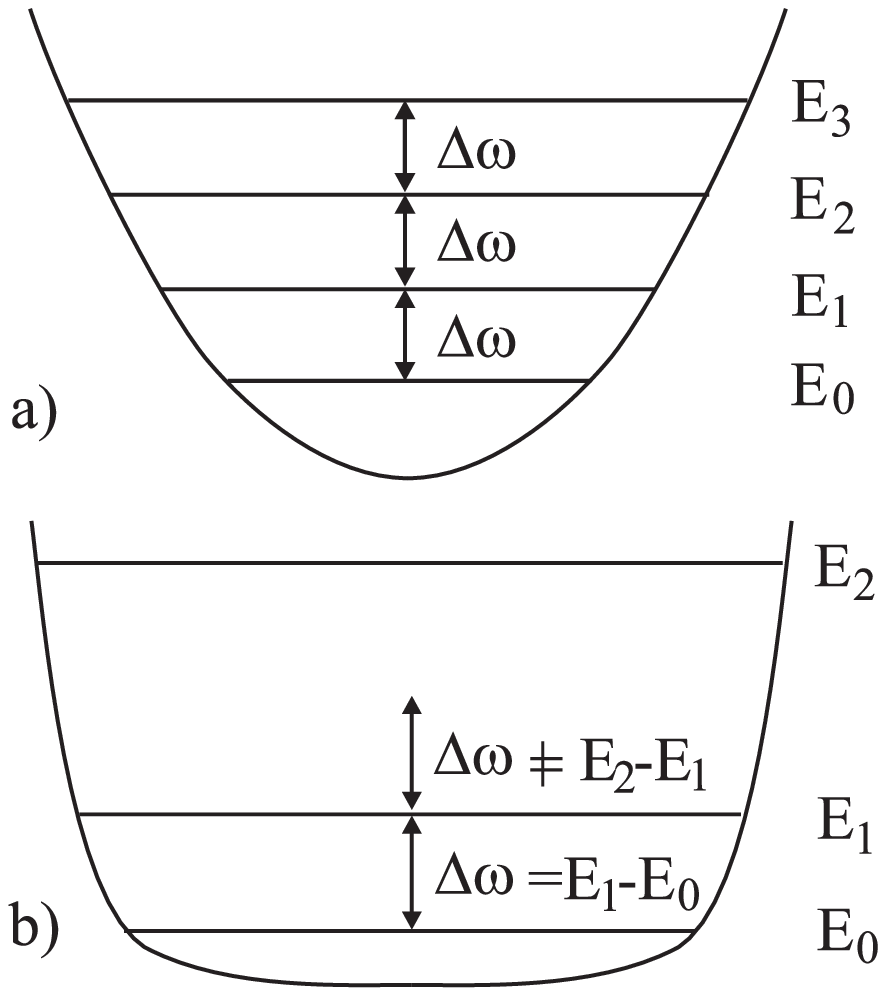}%\vspace{-1cm}
\caption{This diagram shows the ladder of transitions between different trap
energy levels due to the optical potential: a) For a harmonic trap
all transitions are resonant. b) In an anharmonic trap only the first
transition is resonant so that only the first excited vortex state
is populated. }
\end{figure}

$ $

$ $

\begin{figure}[t]
\epsfysize=5.5cm
\hspace{3cm}
\epsffile{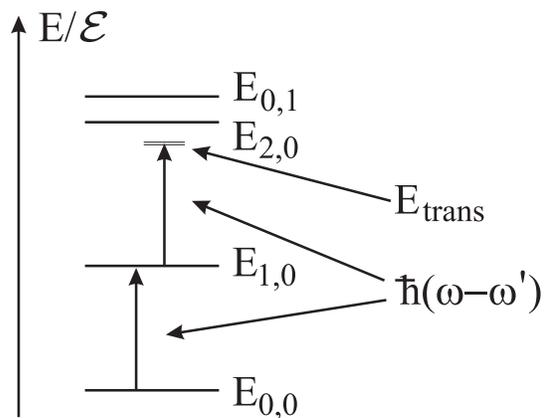}%\vspace{-1cm}
\caption{The lowest lying eigenstates for the anharmonic trap. If the
frequency difference $\omega-\omega^\prime$ is tuned into
resonance with the lowest transition and the optical potential
(\ref{vlvort}) (represented by $E_{\mbox{{\scriptsize trans}}}$) is weak 
transitions to higher states are highly suppressed. }
\end{figure}

$ $

$ $

\begin{figure}[t]
\epsfysize=7cm
\hspace{3cm}
\epsffile{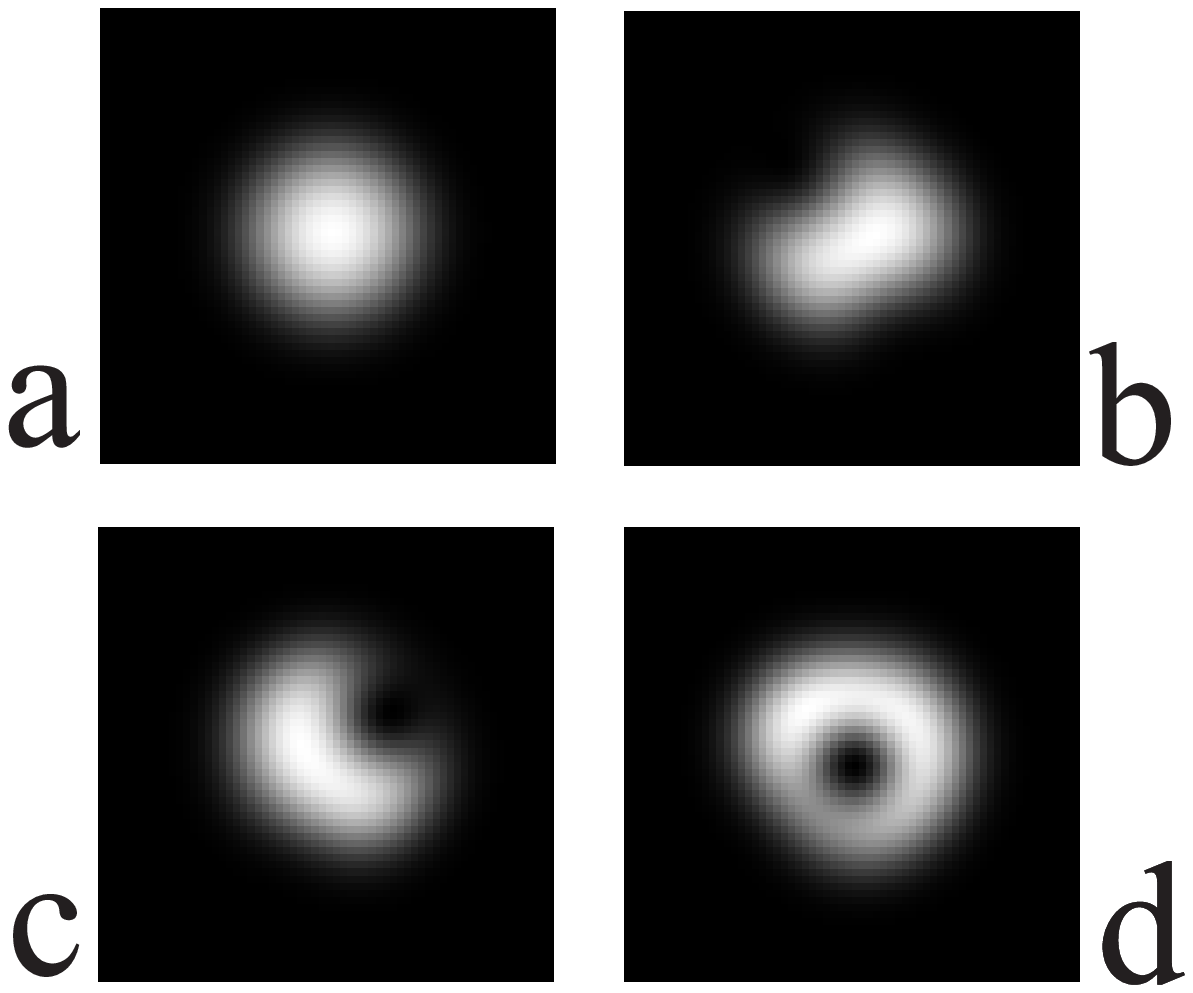}%\vspace{-1cm}
\caption{Numerical simulation of the condensate's
time evolution during the vortex creation. The
modulus squared of the collective wavefunction is shown at different
times (a: 0 ms, b: 87 ms, c: 142 ms, d: 247 ms).  }
\end{figure}

$ $

$ $

\begin{figure}[t]
\epsfysize=7cm
\hspace{3cm}
\epsffile{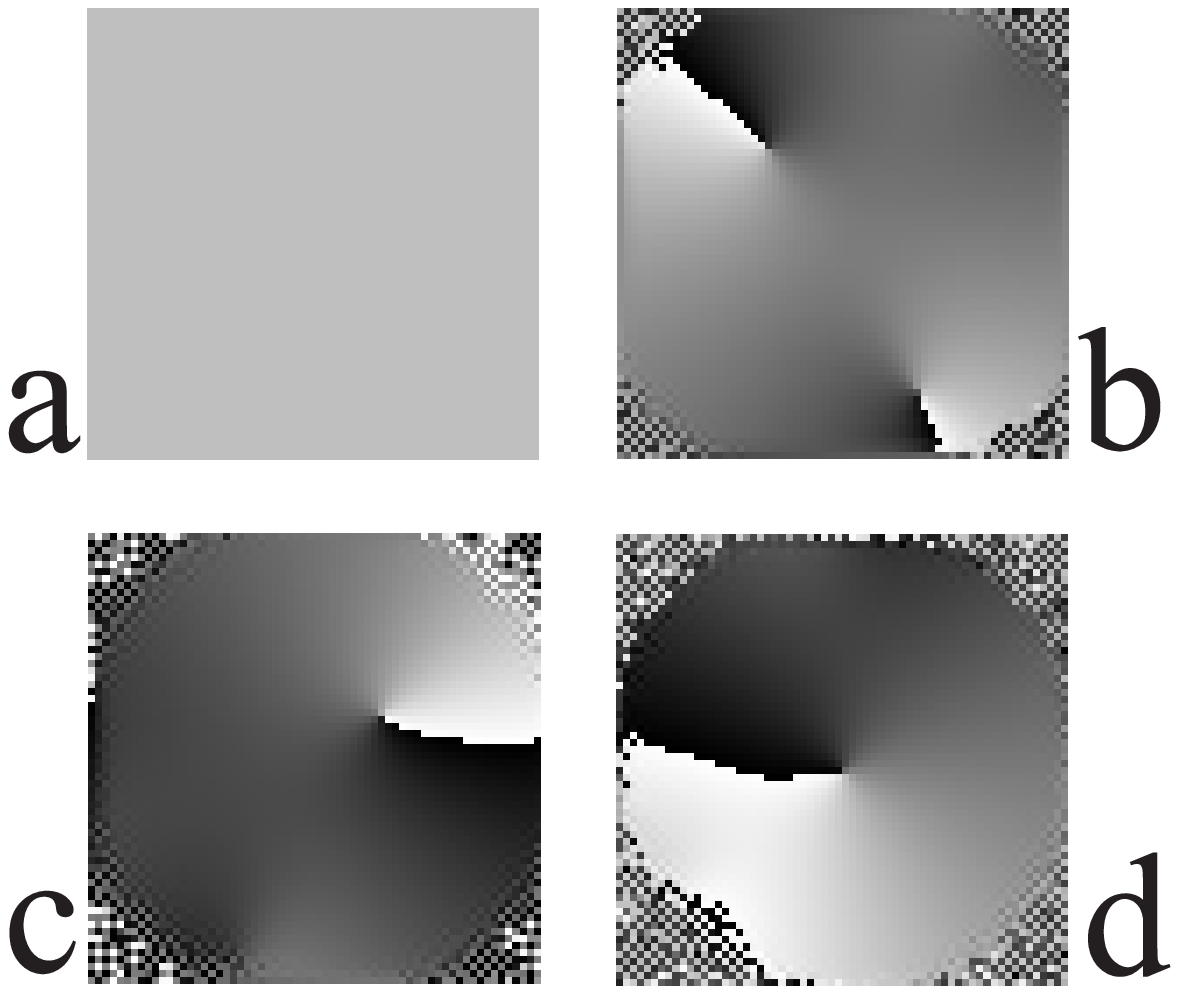}%\vspace{-1cm}
\caption{The same as Fig.~4. The phase of the wavefunction is shown, varying 
between $\pi$ (white) and $-\pi$ (black). A vortex line corresponds
to the end of a sharp borderline between white and black.}
\end{figure}

\end{document}